# EAPESS: An Adaptive Transmission Scheme in Wireless Sensor Networks

Z. Abbas[1], N. Javaid[1], A. Javaid[2], Z. A. Khan[3], M. A. Khan[1], U. Qasim[4]

[1]*CAST, COMSATS Institute of Information Technology, Islamabad, Pakistan*

[2]*Mathematics Dept., COMSATS Institute of Information Technology, Wah Cantt, Pakistan*

[3]*Faculty of Engineering, Dalhousie University, Halifax, Canada*

[4]*University of Alberta, Alberta, Canada*

**Abstract**: - Reduced energy consumption in sensor nodes is one of the major challenges in Wireless Sensor Networks (WSNs) deployment. In this regard, Error Control Coding (ECC) is one of techniques used for energy optimization in WSNs. Similarly, critical distance is another term being used for energy efficiency, when used with ECC provides better results of energy saving. In this paper three different critical distance values are used against different coding gains for sake of energy saving. If distance lies below critical distance values then particular encoders are selected with respect to their particular coding gains. Coding gains are used for critical distances estimation of all encoders. This adaptive encoder and transmit power selection scheme with respect to their coding gain results in a significant energy saving in WSNs environment. Simulations provide better results of energy saving achieved by using this adaptive scheme.

1. INTRODUCTION

Energy consumption in WSNs is an important issue. As sensor nodes are composed of Computation unit and Radio unit, then both of them are main sources of energy consumption. Various strategies have been employed for energy optimization like switching transceiver of radio unit to sleep mode that, reduces energy consumption with respect to active mode. Strategy for switching transceiver is to activate transmitter only when there is data to send while, receiver is kept into awake mode for reception of data that is not an energy efficient method. Solution is to keep receiver in sleep mode and activate it for some time, if no activity is detected then go back into sleep mode. Clustering was also an implemented solution to lower energy consumption.

However, these solutions may cause channel impairments (that can cause loss of data) which, on other hand requires retransmissions. Automatic Repeat Request (ARQ); is a technique to overcome this problem. In this regard, Forward Error Correction (FEC); was used to overcome this problem that reduces number of retransmissions and frame error rate. Similarly, Error Control Coding (ECC); was another technique to lower energy consumption in which, by lowering transmit power link reliability was increased. Stronger codes provide better performance with low transmit power as compared to simpler codes. In Previous work for energy efficient transmission Reed Solomon (RS) codes and Convolution codes were used. To acquire a required Bit Error Rate (BER) higher rate encoders were used. Decoding of convolutionally encoded data was performed with Vterbi algorithm using hard decision decoding or soft decision decoding.

In this paper, we investigated three different coding schemes like RS, CC-Hard and CC-Soft. These techniques are discussed with respect to coding gain and critical distance values. Critical distance $D_{CR}$ estimation is performed for each coding gain and against each critical distance particular transmit power and encoder is used. In our work, three different cases have been discussed 1st includes critical distance values with respect to RS code, 2nd includes critical distance value with respect to CC-Hard code and lastly critical distance value for CC-Soft is calculated. Then with respect to these three critical distance values particular transmit powers and encoders are selected.

Section 2 of this paper includes related work and motivation, Section 3 explores transmit power estimation and ECC techniques used for energy saving. Performed simulation results have been discussed in section 4. Section 5 concludes our work.

## 2. RELATED WORK AND MOTIVATION

After analyzing ECC Sheryl *et al.* [1] estimated minimum transmit power for energy efficient transmission of data in different environments like free space model and far field region. Then, analyzed bit error rate for different values of $E_b/N_o$ with different encoding schemes. He also analyzed decoder power consumption when data is reached to its destination. Then lastly, he calculated energy savings achieved from different ECC techniques.

Pellenz *et al.* [2] in his work focused on radio channel models, that may experience different channel impairments, these impairments can effect data being transmitted, and if it happens so retransmissions of that data is required that consumes extra energy on other hand. So, in his work error control strategies and error correcting codes have been used to avoid this thing. He investigated tradeoff between transmission energy consumption and processing energy consumption using convolutional codes. His work enabled appropriate selection of encoder complexity in order to improve network life time. At the same time work of Atta et al. [3] was also based on adaptively selection of code rate in an Orthogonal Frequency Division Multiplexing (OFDM) system.

Sonali *et al.* [4] discussed that how energy is affected by ECC and modulation type used. She discussed utilization of ECC from energy saving perspective. She discussed that error correcting capability and code word length of ECC and modulation type are of great importance in sense of energy consumption. So, optimal ECC and modulation scheme is used for better energy saving. Sapna et al. [5] discussed a method to correct errors based on Euclidean codes. On the basis of minimum Euclidean weight errors are corrected in transmitted data.

To prolong network life time Meghji *et al.* [6] used multi-hop concept with transmit power control in WSNs. He tested transmit power control in multi-hop networks to check energy consumption results. Then showed that energy using short range multi-hop concept cannot be saved.

Transmit power for WSNs discussed by Lavratti *et al.* [7] is adaption of minimum transmits power for extension of network life time, and to guarantee that transmitted data has been successfully received. He presented Transmission Power Self Optimization (TPSO) technique, by providing greater amount of efficiency, whether it's a noisy medium.

Due to constraints of fault tolerance, Scalability and cost and power consumption Akyldiz *et al.* [8] presented that new techniques are required to overcome these problems over different layers by presenting several projects that are being performed in WSNs.

Sandra *et al.* [9] provided a survey for main causes of energy loss in WSNs. They considered that most of loss occurs because of electronic circuitry then they provided a comparison between different MAC and routing protocols that have been designed to overcome power consumption problem and by providing efficient and reliable data transfer.

Work of Pranali *et al.* [10] is based on 802.16e to enhance power efficiency by employing power saving strategies used for standard sleep mode operation. They proposed an algorithm to save power consumption and mean delay of power saving mechanism for sleep mode operation.

From looking all above discussed works we proposed our adaptive transmit power and encoder selection scheme for energy saving in WSNs, that is somehow a major issue in sensor networks domain.

3. **TRANSMIT POWER AND ERROR CONTROL CODING**

Concept of transmit power is used with respect to distance, less distance requires less transmit power while large distance requires greater transmit power. Similarly, Error Control Codes are used for efficient data transmission, Block Codes or Convolutional Codes are types of them.

A. Transmit Power

Signal transmitted with minimum power is an efficient source of energy minimization. Receiver has to maintain a minimum signal to noise power called $E_b/N_0$ for successful operation. Transmit power required to transmit a signal represented by $P_{TX}$ in free space model is given by [1]

$$P_{TX} = \frac{S}{N} N \left(\frac{4\pi d}{\lambda}\right)^2 \tag{1}$$

$$P_{TX} = \eta \frac{E_b}{N_o} mKTB \left(\frac{4\pi d}{\lambda}\right)^2 \tag{2}$$

Here, $E_b$ is minimum energy that is required for one bit where $N_0$ is noise power spectral density. $\eta$ is spectral efficiency expressed as information rate to bandwidth ratio, $m$ noise proportionality constant, $K$ Boltzman constant and $T$ absolute temperature all together represents signal noise $\lambda$ represents transmitted wavelength at some distance represented by $d$ in free space model. Transmitted power of signal depends on distance between transmitter and receiver. If it is short then less transmit power is required and if it's large then signal has to be transmitted with maximum power. When distance is greater than sometimes data needs to be retransmitted by using some ECC. So that, actual data should be sent and received with in time.

B. ECC

In ECC extra bits are added in information for receiving exact information that was originally transmitted. For example, information that is being sent is *u* with length *k* then extra parity bits are added to information *u* to form a codeword *x*, these extra redundant bits enables decoder to correctly decode *x* received bits. So, over noisy channel where there are maximum errors occurring chances there ECC provides a better Bit Error Rate (BER) at given Signal to Noise Ratio (SNR) value. Difference between SNR levels to reach a certain BER value in coded and un-coded system is called coding gain. Here, in our paper we used three different types of encoders based on coding gain of these encoders. Reed Solomon Encoder (RS), Convolutional Encoder with Hard Decision Decoding and Convolutional Encoder with Soft Decision Decoding are these encoders. RS codes are block based error correcting codes, where information after adding redundant bits is sent in form of blocks. For example, there are *k* data symbols of *s* bits to make *n* symbol codeword. Therefore, *n-k* parity symbols of each *s* bit, and decoder for RS encoding can correct up to *t* symbols that contain errors in codeword.

Convolutional codes are frequently used to correct errors in noisy channels and have good error correcting capability over bad channels. Threshold Decoding, Sequential Decoding and Viterbi decoding are decoding methods for convolutionally encoded data. Here, Viterbi with Hard and Soft decision decoding is used. In Hard Decision decoding is performed when received sequence is digitized 1st then decoding is performed, so it makes an early decision without considering a bit is 0 or 1. While, in Soft Decision decoding is performed before digitizing received data. In our case, selection of encoders is performed on the basis of distance, depending on distance encoders are used adaptively for better results energy savings.

Algorithm 1

C. Energy Saving

Energy required to transmit data equals to require transmit power divided by total information transmission rate $R$ that is being transmitted. Energy required to transmit un-coded data and that for coded data is calculated from given formulas. Where coded data transmission requires ECC gain values for energy calculation [1].

$$E_{TX,U} = \frac{P_{TX,U}}{R} \tag{3}$$

$$E_{TX,ECC} = \frac{P_{TX,ECC}}{R} \tag{4}$$

Eq. (5) below gives energy saving from either coded or un-coded data transmission.

$$\Delta ES = E_{TX,U} - E_{TX,ECC} \tag{5}$$

Using above eq. energy saving for all encoders used is calculated for 100m × 100m area. Here, energy saving has been calculated for three different cases using three different encoders with their different coding gains, and with critical distance limit of all coding gains. Energy saving from all three cases have been calculated and presented in simulations section.

## 4. SIMULATIONS

This section provides details of concerning simulation environment. Our proposed scheme's performance is compared with already implemented techniques and MATLAB was used simulation tool. In simulation environment 500 sensors are randomly deployed in a $100m$ x $100m$ area and sink is placed in corner of that area. Firstly, transmit power and encoder selection is performed, then energy saving from all cases is calculated from our all implemented encoding techniques, Table. I provides information about parameters used in simulations.

Table 1

Figure 1

### A. Results and Discussions

Simulations are performed regarding critical distance values for all encoders. In all cases when distance between sensor node and sink is less than critical distance of low coding gain encoder then low coding gain encoder is used if it is greater than medium coding gain encoder is used and when distance is large enough then higher coding gain encoder is used.

As energy saving is calculated from energy spent for un-coded data transmission to energy spent for coded transmission. Therefore, power required for transmission of un-coded data is presented in Figure1. It shows that power needed for transmission is far greater because of no coding scheme or encoder is used, that's why a lot of power is required for transmission of data if sensor nodes are distant from each other. Reason behind this becomes clear as sensor nodes in sensor network are deployed randomly and their transmit power is estimated during increasing number of rounds. In one round sensor node may be nearer to sink while same sensor node in next round may be farther from sink, same is happening in Figure 1 in first few rounds when sensor is placed nearer to sink less transmit power is required while, in some rounds larger transmit power is required which means that sensor has moved to some farthest place.

Figure 2

For case 1 where distance is less than critical distance of low coding gain encoder then here RS encoder is used. Transmit power for encoder is estimated for transmission of data from sender to receiver. Same happens in case 1 as transmit power for RS encoder is being estimated then location of sensor not remains same over 100 rounds, as it keeps on changing its location because of its random placement in network. Sometimes, because of encoder type and placement near to sink sensor node requires less power in some rounds while, in some rounds it needs higher power for transmission. Figure 2 shows changing transmit power behavior of sensor over 100 rounds in sensor network.

Figure 3

Figure 4

Figure 3 shows energy saving after transmission of data from sensor node using RS encoder over 100 rounds. As number of rounds is being increased energy saving also increases and becomes constant after approaching some certain rounds. Energy saving depends on transmit power and type of encoder being used, as RS encoder is used for distances below critical distance value energy is being saved and if critical distance value is reached, then encoder with greater coding gain is used to transmit data which saves energy by efficient transmission. Energy saving increases during first 30 rounds and after these rounds energy saving becomes constant, whatever, sensor node location is energy saving remains same.

Figure 5

Figure 6

For the case where convolutional encoder with hard decision decoding is used there, critical distance value for that encoder is estimated, if distance values are less than estimated critical distance value then (CC-Hard) encoder is used. If distance values are greater than critical distance values then Convolutional encoder with soft decision decoding (CC-Soft) is used for efficient data transmission. Figure 4 shows an estimated transmit power for CC-Hard encoder. An abrupt changing behavior of transmit power is shown, as it is depending on critical distance value so during each round distance is changing randomly because of sensor movement. That's why sensors during most of rounds require higher transmit power for data transmission.

After transmit power estimation, starts process of sending data by using selected encoder, and if distance between sensor and sink is below critical distance value then CC-Hard is used for data transmission if critical distance value is reached then CC-Soft encoder is used for sake of energy saving of that sensor node. Figure 5 shows somehow, same behavior for energy saving as for RS encoder however; this behavior during some rounds becomes same. Reasons is that because of random placement and higher transmit power values during most of rounds, less energy saving occurs. And saving obtained from this type of encoder and RS encoder becomes nearly same after 40-50 rounds, while, saving before 40 rounds is greater for this type of encoder.

Figure 6

Figure 6 provides estimated transmit power for selected encoder that is (CC-Soft). Figure shows little bit different behavior for transmit power from others, because of CCS encoder having a greater coding gain with respect to others. Simulation is performed for 100 rounds to estimate exact transmit power for efficient transmission below critical distance value. And for values above critical distance RS Encoder is used for energy efficient transmission. In figure transmit power behavior for sensor keeps on changing during increasing number of rounds because of movement of sensor node during each round. During, first few rounds sensor needs less transmit power then after them it requires higher transmit power values. All this happens because of random placement of nodes and distance values that sometimes becomes less and sometimes greater.

Figure 7

Energy saving after using selected encoders is presented in Figure 7. With increasing number of rounds energy saving reaches to its maximum value and finally becomes constant for remaining rounds. Using this case large amount of energy is saved as compared to other encoders, it is possible because of required transmit power for sensors that on other hand is depending on distance and encoder type being used. Here, in this case an efficient encoder is used if sensor keeps on changing its location over 100 rounds then use of this encoder saves energy. However, increase in energy saving occurs during first few rounds, then after few rounds energy saving like others becomes constant.

Figure 8

Figure 8 shows a comparison between our adaptive scheme and default schemes used. In all of three cases CC-Soft shows best results of energy saving in our adaptive scheme as compared to default CC-Soft, then adaptive CC-Hard and RS encoders shows good results for energy saving. Default schemes show a constant behavior for energy saving over varying distances. In figure a clear view comparison of energy saving is presented for adaptively selected encoders over 100 rounds to those encoders that are already being used. Energy saving from default encoders increases during first few rounds then it shows a constant behavior over changing rounds. CC-S default shows greater amount of energy saving when compared to other default encoders, while CC-S adaptive shows a great amount of energy saving as compared to other adaptive encoders as shown. Energy saving from our proposed scheme is because of critical distance at which decoder energy consumption becomes equal to energy saving at encoder side.

So, our adaptive technique shows an energy efficient behavior as compared to default encoding schemes. Among them CC-S adaptive because of its efficient coding gain performs best in terms of energy saving, then RS and CC-H adaptive performs best.

## 5. CONCLUSION

This paper presents the efficient encoder selection and transmits power with respect to its critical distance results in energy saving in WSNs. Encoder selection is performed by using critical distance which is estimated from coding gain of that encoder. Transmit power is estimated for selected encoder to transmit data efficiently. ECC in this context, becomes energy efficient as encoders and their transmit powers are selected adaptively, that results in energy saving to these particular encoders.

# Tables

Table 1. Simulation Parameters

| Parameters | Value |
|---|---|
| Standard | 802.15.4 |
| Frequency band | 2.45 $GHz$ |
| Data rate | $250 Kbps$ |
| Distance | $1.0m - 100m$ |
| Receiver Noise Figure | 5 $dB$ |
| $E_b/N_0$ | 6.76 $dB$ |
| $Eta(\eta)$ | 0.0030 |
| SNR | 0.0202 |

**Algorithm 1 :-** Proposed Algorithm for RS, CC-Hard and CC-Soft

1: $D \leftarrow RandomDistance$

2: $D_{CR_{RS}} \leftarrow Critical Distance of RS$

3: $D_{CR_{CCH}} \leftarrow Critical\ distance\ of\ CCH$

4: $D_{CR_{CCS}} \leftarrow Critical\ distance\ of\ CCS$

5: $P \leftarrow Reed\ Solomon\ Encoder$

6: $Q \leftarrow CC\ Hard\ Encoder$

7: $R \leftarrow CC\ Soft\ Encoder$

8: $P_{TX} \leftarrow TransmitPower$

9: $CASE\ \ 1$

10: **if** $D \leq D_{CR_{RS}}$ **then**

11: $\quad Estimate\ P_{TX}$

12: $\quad then\ use\ P$

13: **end if**

14: **if** $D \geq D_{CR_{RS}}$ **then**

15: $\quad Estimate\ P_{TX}$

16: $\quad and\ use\ Q$

17: **end if**

18: $CASE\ \ 2$

19: **if** $D \leq D_{CR_{CCH}}$ **then**

20: $\quad Estimate\ P_{TX}$

21: $\quad and\ use\ Q$

22: **end if**

23: **if** $D \geq D_{CR_{CCH}}$ **then**

24:     $Estimate\ P_{TX}$

25:     $and\ use\ R$

26: **end if**

27: $CASE\ \ 3$

28: **If** $D \leq D_{CR_{CCS}}$ **then**

29:     $Estimate\ P_{TX}$

30:     $and\ use\ R$

31: **end**

32: **if** $D \leq D_{CR_{CCS}}$ **then**

33:     $Increase\ P_{TX}$

34:     $and\ use\ R$

35: **end if**

# Figures

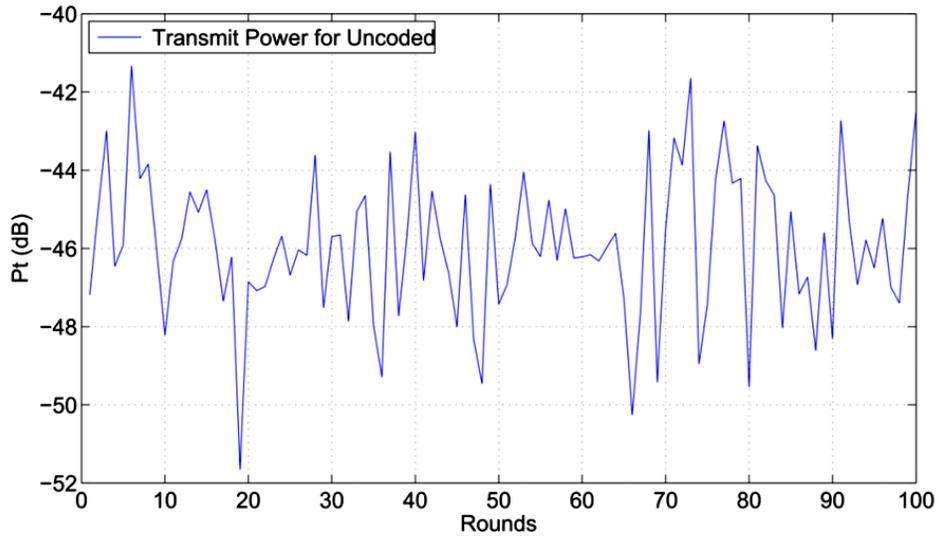

Figure 1. Estimated Transmit Power (dB) for Un-coded Transmission

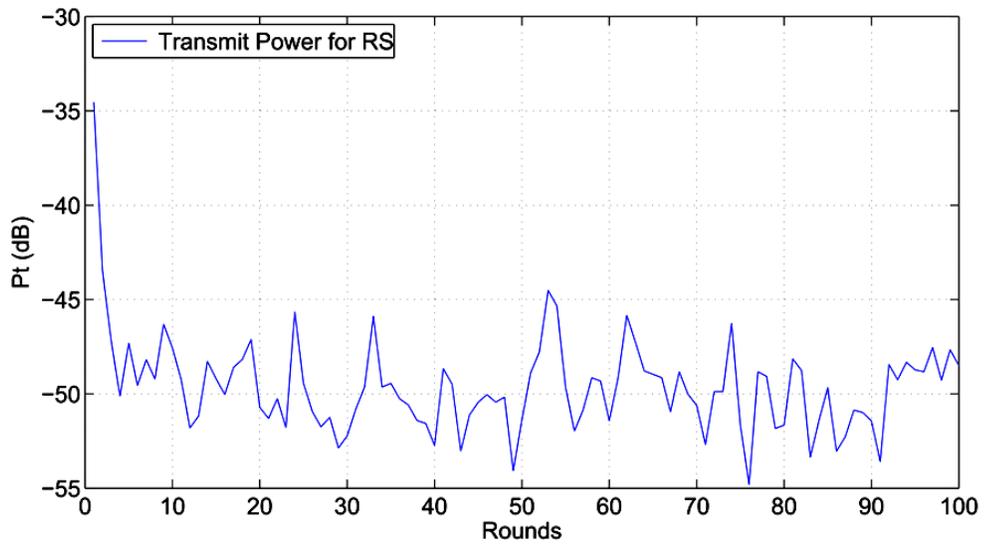

Figure 2. Estimated Transmit Power (dB) for RS Encoded Transmission

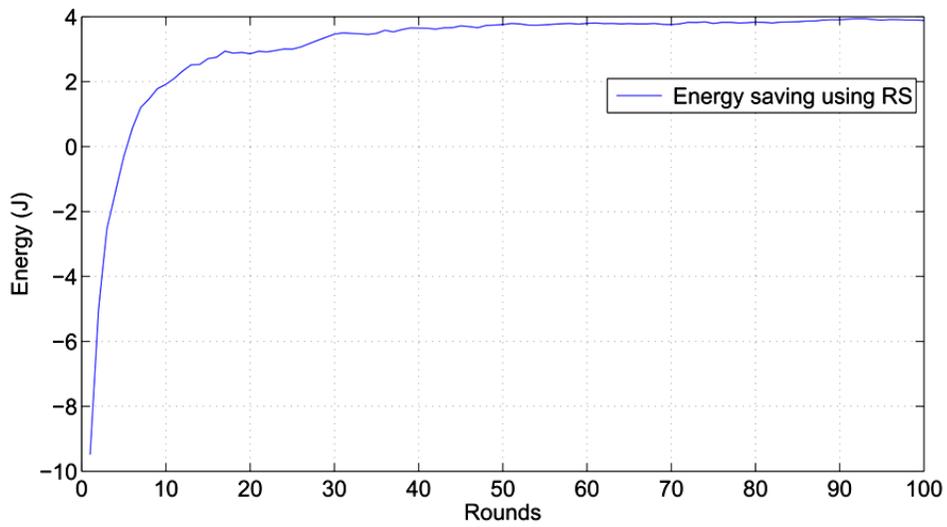

Figure 3. Energy Saving for RS Encoded Transmission

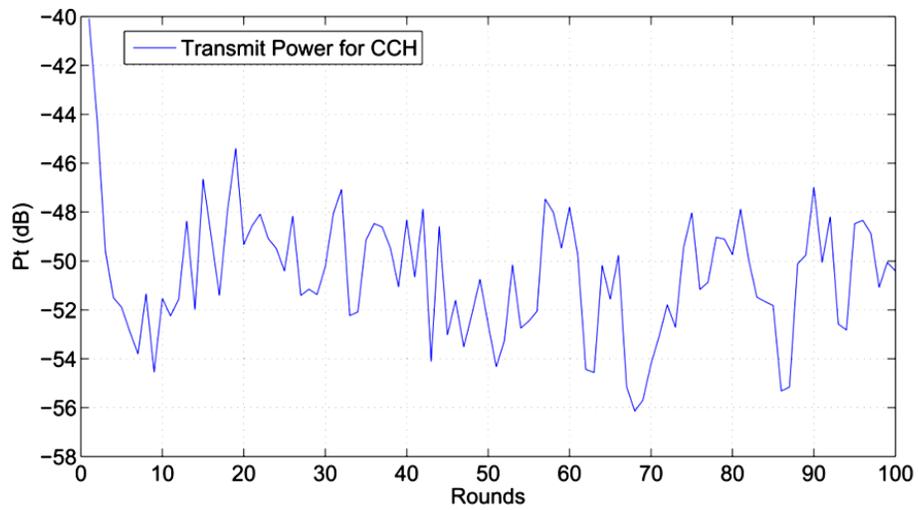

Figure 4. Estimated Transmit Power (dB) for CCH Encoded Transmission

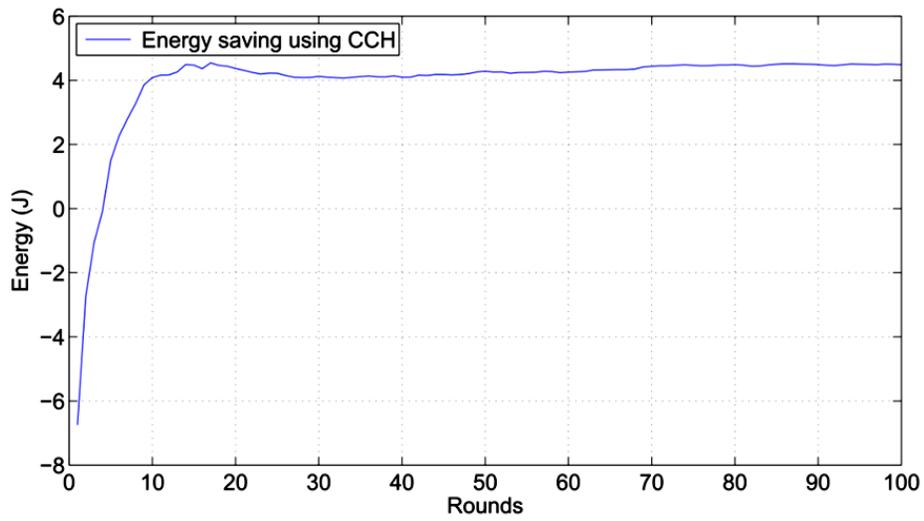

Figure 5. Energy Saving for CCH Encoded Transmission

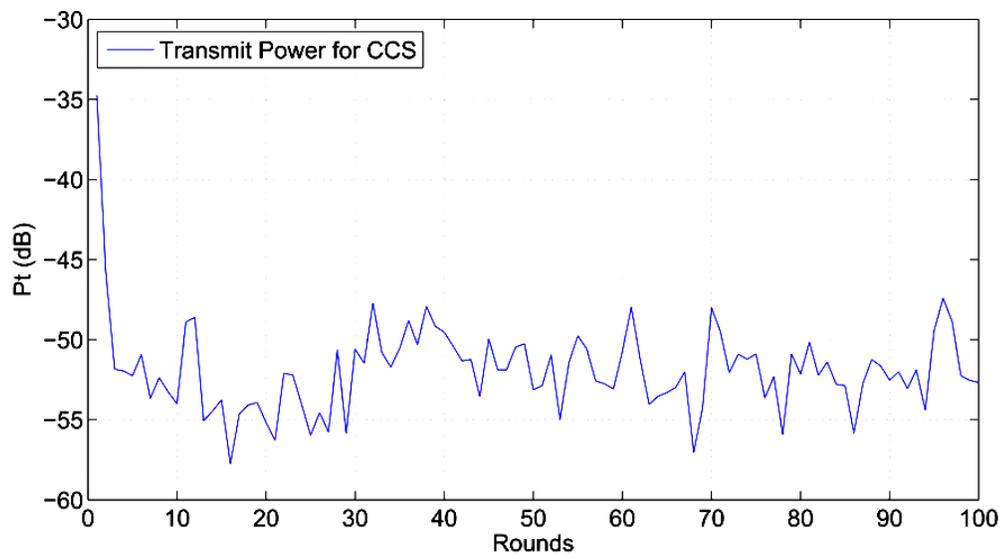

Figure 6. Estimated Transmit Power (dB) for CCS Encoded Transmission

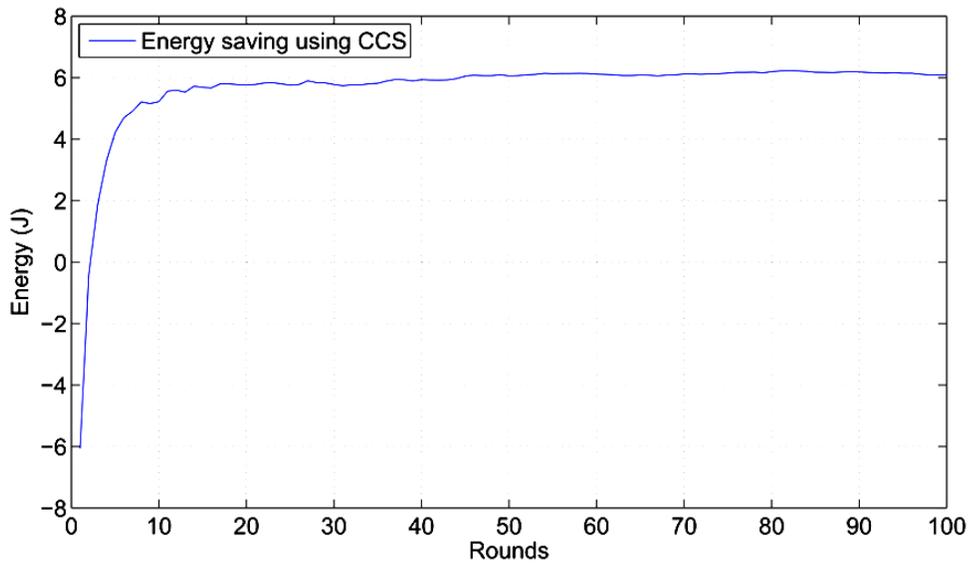

Figure 7. Energy Saving for CCS Encoded Transmission

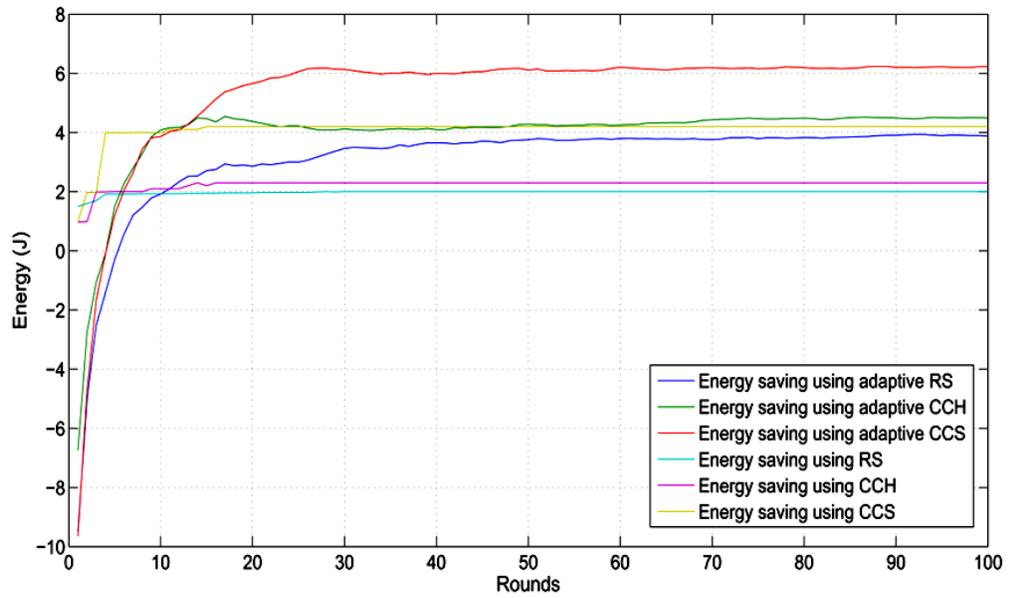

Figure 8. Energy Saving Comparison of Adaptive and Default Scheme